# Gender Issues & Information Communication Technology for Development (ICT4D): Prospects and Challenges for Women in Nigeria

Kwetishe Joro Danjuma[1], Bayo Mohammed Onimode[2] and Ochedikwu Jonah Onche[3]

[1] Department of Computer Science, Modibbo Adama University of Technology,
Yola, Adamawa State, Nigeria

[2] Department of Computer Science, Federal University of Technology,
Minna, Niger State, Nigeria

[3] Computing Department, University of East London,
London, United Kingdom

**Abstract**

Information and Communication Technology is a compendium of interrelated applications, products and services that can either deepen or alter gender equality. It has successfully transformed education, businesses, healthcare, entertainment, politics and good governance within the Global North; providing equitable access to developmental framework driven by ICTs. However, in drafting the core developmental objectives of sustainable development, there has been considerable gender digital divide limiting women access to resources based on their gender, ethnicity, socio-cultural bias and the rights to utilize such resources for development. In realization of the United Nation Millennium Development Goals within the Global South specifically Nigeria, women are often marginalized or excluded from ICT policy drafting and imbalance associated with ICT. This paper identifies and evaluates gender issues and information communication technology, with focus on the challenges and prospects for women empowerment in Nigeria. The study critically examined research literatures and conducted research survey on the prospects and challenges of promoting gender equality and women empowerment through ICTs; and identifies policy implication for Nigeria. The research survey used a random sampling technique with a target sample size of eighty respondents. Data gathered from the questionnaire was analyzed using Statistical Package for Social Science version 19, and the result was presented using ANOVA, and descriptive analysis. The study reveal gender inclusiveness in policy drafting as a key driver for socio-economic development, improved healthcare and women empowerment in Nigeria. We recommend a deliberate ICT policy that attract and encourages women participation in ICT developmental framework.

## 1. Introduction

Globally, humans are experiencing an extraordinary transitioning. While increased technology adoption has helped many people in their daily lives, for some groups, digital disparities still remains and this is attributable to their lack of technical skills. There is need for the acquisition of basic skills that are needed to use the technologies effectively and safely. It is also important to have a clear understanding of the needs of the target groups [1]. According to the ITU World Telecommunication/ ICT Indicators database, more men than women use the internet: globally, 37% of all women are online, compared with 41% of all men. This corresponds to 1.3billion women and 1.5billion men. The developing world, Nigeria inclusive is home to about 826million female internet users and 980million male internet users while the developed world is home to about 475million female internet users and 483million male internet users. Thus the gender gap is more pronounced in the developing world, where 16% fewer women than the men use the internet, compared with only 2% fewer women than the men in the developed world [2].

The MDG Goal 3 is: To promote gender equality and empower women. This should commence with the elimination of gender disparity in primary and secondary education. In the developing regions of the world, the gender parity index (GPI) is defined as girls' school enrolment ratio in relation to boys' enrolment ratio at every level of learning, and this is in the range of 0.97 and 1.03 which is in the accepted measure for parity [3]. Conversely, a closer look reveals a major gender disparity among world regions at all the levels of education. UN Women, the United Nations entity for gender equality and the empowerment of women – reports that Gender-based inequalities in decision-making influence still persist. Gender discrepancies are more obvious at the secondary education level. Girls continue to be at a disadvantage to boys in sub-Saharan Africa, Western Asia and Southern Asia. This situation is most extreme in sub-Saharan Africa, where the gender gap has really widened, with the GPI falling from 0.66 to 0.61 between 2000 and 2011 [4]. Whether in the private or public area, from the top levels of governance/decision-making to the family units, women continue to be denied the equal opportunity with men to participate in decisions, even that which affect their lives [5].

There is the need to share information on policy and institutional issues, skills development, exchange/transfer and adaptation of new technologies; this will potentially attract financing or investment from private and other non-state sources as well as the public. The UN Secretary-General Ban Ki-moon has acknowledged, "Equality for women and girls is not only a basic human right it is a social and economic imperative. Where women are educated and empowered, economies are more productive and strong. Where women are fully represented, societies are more peaceful and stable." Thus the ICT sector remains a buoyant and growing sector for employment, and a key economic factor underpinning both national and international development [3]. A number of works have been carried out in the area of Gender and ICT, probing the position that ICTs plays in tackling gender disparities in education and literacy in developing countries [6-7]. Though, limited research exists on how the level of education influence ICT perception and use among men and women, or on how this, in turn, affects gender standards and relations in an organization or work place.

This paper is written with the major question as: what are the prospects and challenges for women in Nigeria as regards gender equality and access to ICTs and how this control the ways that ICTs are perceived and used in Nigeria compared to that of their male counterpart? We also examined gender, education, and ICTs, as they shape the basis for the study. This is followed by a synopsis of the research design, a sketch of the cases being studied, and a presentation and discussion of the analysis. The work is concluded with a summary of key results and the propositions for future research in this area.

## 2. Background and Related Work

Information and Communication Technologies (ICTs) comprise a complex, heterogeneous and interrelated set of goods, applications and services used to produce, process, distribute and transform information [8]. ICT tools have helped people find, explore, analyze, exchange, and present information without discrimination; and can provide quick access to ideas and experiences from a wide range of people, communities, and cultures [9]. ICTs have created new economic and social opportunities all over the world, and women have been able to use ICTs to support new forms of information exchange, organization, and empowerment [10]. ICT can empower women and girls if the existing gender disparities such as lack of infrastructure, education and skills, financial resources, power and decision making, social and cultural bias associated with ICT utilization among Nigeria women are identified and eliminated [11]. ICT can empower rural women economically by reducing trade distortions, eliminate poverty, empower weaker segments, increased productivity, get information about market prices, etc. It is however, possible only if a nation follows sound ICT strategies and policies [12].

An analysis of the key challenges to the utilization of ICT to promote socio-economic development, and in particular, the achievement of the Millennium Development Goals (MDGs) with specific focus on poverty alleviation, health, education, gender equality, and environmental sustainability was conducted [13] to further illustrate the possible impacts of ICT as an enabling tool for socio-economic development that directly or indirectly impact on the achievements of the MDGs. Technical and socio-cultural barriers exacerbate women's exclusion from the benefits of ICT for development [14]. Inaccessibility to ICT has widened educational, healthcare, information and communication gaps between urban and remote communities [15]. In addition to the identified challenges, [14] identified location, lack of infrastructure and connectivity, time and money, lack of relevant content, low education and literacy, social norms and perceptions as women most significant barriers to the utilization of ICT for development (ICT4D). The systematic documentation of information and communication needs of women in the global south [16], assert that obstacle to ICT accessibility and utilization are generally structural (time, location, and illiteracy) and not personal (for example, a prohibition from a relative). These challenges greatly account for the reasons why Nigerian women are often underrepresented in terms of ICT access and utilization despite the great emphasis placed on the use of ICT in Nigeria [17].

The new information and communication technologies (ICTs) such as mobile phones and the internet are considered [18] important instruments for advancing socio-economic development throughout the world. According to [19], ICTs as a new form of technology are socially deterministic, with varied implications for women in terms of employment and empowerment that is dependent on the context within which the ICTs are utilized. Women's empowerment and inequality are inseparable and necessary components for the realization of sustainable socio-economic development [20]. Gender is considered a cross-cutting theme of the MDGs and the fifth goal calls for gender equality, while recognizing ICT as key driver to promoting women empowerment; and the women empowerment is also thematic as a policy initiative to enhancing development [21]. Gender equality entail all human beings, both men and women are free to develop their personal abilities and make choices without the limitations set by stereotypes, rigid gender roles and prejudices [22]. The United Nations Secretary-General, Kofi Annan affirms that the so-called digital divide is actually several gaps in one: There is a technological divide-great gap in infrastructure. There is a content divide. A lot of Web-based information is simply not relevant to the real needs of people. There is a gender divide, with women and girls enjoying less access to information technology than men and boys [23]. A study funded by the UK Department of International Development provided a comprehensive view of gender in Nigeria [24], and assessed progress made in key areas including employment and livelihood, education and health, political representation, and domestic violence against women (VAW). In their findings, women and girls were found to have suffered systematic disadvantage and discrimination that is magnified for those in so-called poorest states and sectors of the Nigeria society. Despite the so-called gender digital divide, the paucity of statistical data on gender digital divide makes it difficult, if not impossible for inclusion in ICT policies, plans, and strategies [25].

However, an effective ICT policy will empower and protect the rights of the girl child, increase women socio-economic opportunities and contribute to ICT innovation and research; and promote broader equity with emphasis on providing adequate access to democratization and political resources, information and education tool [11]. According to [25], education remains the single most important factor in improving the ability of girls and women in developing countries to take full advantage of the opportunities offered by information technology. There is also the need to change women's attitude towards the use of ICTs, and help them overcome technophobia by deliberate government ICT policies that provide and encourage educational and economic empowerment to women [17]. In line with education, [14] asserts that ICTs and public access to computer with internet connectivity can assist community development efforts and in turn bridged the so-called gender digital divide. The gender differences in education accessibility and constraints to gender equality in Nigeria were appraised [26], and illiteracy was identified as symptom of exploitation, inequality and poverty; and if not appropriately addressed would have adverse effects on manpower development in Nigeria. In a study of illiteracy rate and accessibility to telephone facilities, computers, and internet facilities in Africa, gender digital divide was discovered between those with access to the internet and related technologies and those who do not [27]. [20] Opined that high female illiteracy, patriarchal societies, and socio-cultural barriers can perpetuate gender inequality and the efficacy of multimillion dollars mobile phone investment programs to support women empowerment. In relation to patriarchal society, [28] examined the conceptual and material bases of patriarchy and gender inequality in Nigeria; identified dimensions of gender inequality and discrimination; and opined socio-cultural and political factors leads to gender discrimination.

The benefits of ICTs access and utilization are both individual and collective [14]: the individual benefits include increased self-esteem, reduced isolation, access to markets, empowerment and access to health information; and collective benefits include economic growth, improved healthcare and education, capacity building and cultural transformation. According to [29], ICTs have to be locally appropriated by women in order to facilitate their empowerment; and build on existing socio-economic and organizational community structure so they can lead not only to the individual woman, but the collective empowerment of women. In order to ensure the entire population reap the many benefits of ICTs, [16] opined a clear understanding of the specific needs of women and other disadvantage group is imperative. However, the many benefits of ICTs cannot be effective if it does not address women's access to and control over technology in relation to the opportunities that abound to use the resources in their context [30]. More so, there are some 162.5 million people in Nigeria, 49% are female; some 80.2 million girls and women; and investment consideration in women and girls will increase productivity, promote sustainable growth, peace and better healthcare [24]; and in turn foster development.

## 3. Methodology

In order to study Gender Issues and ICT4D in addressing the challenges and prospects of women and girls empowerment in Nigeria; a Gender Issues ICT4D survey was designed and distributed among individuals living in Nigeria without gender bias. The participants considered were aged 18 and above, working in either in government, private establishments, self-employed/unemployed, or student sector. In order to ensure research findings reliability, and a high precision estimates without bias, an Internet Protocol address was used to restrict single response per system.

### 3.1 Research Instrument

The research instrument used in this research was based on a gender survey questionnaire adapted from Gender Evaluation Methodology [31]. The GEM questionnaire was structured into three sections. The first section comprised of demographic details of respondents including gender, profession and age. The second section consists of ICT usage habits of the respondents. The third section consists of ICT evaluation and competencies of women and girls in ICT access and utilization. The research question was a 5-point Likert scale on ICT knowledge, knowledge and gender inclusiveness.

### 3.2 Data Collection

The data for this research were gathered from structured survey questionnaires that were started on 07-24 July 2014. A total of 112 participants responded to the online survey questionnaires, out of which 80 participants responded to the survey in full, and 32 responses were partial and therefore discarded due to incomplete filling of the questionnaire. The 80 respondents were completed and used for analysis. The 80 completed respondents account for 89.6% return rate for the questionnaires administered.

### 3.3 Data Analysis

The 89.6% return rate was coded in Statistical Package for Social Sciences (SPSS) and analyzed using descriptive statistics and the Analysis of Variance (ANOVA). Descriptive statistics (frequency, ANOVA, percentage, means, and standard deviation) were employed in the analysis of data using SPSS version 19.

## 4. Results and Discussion

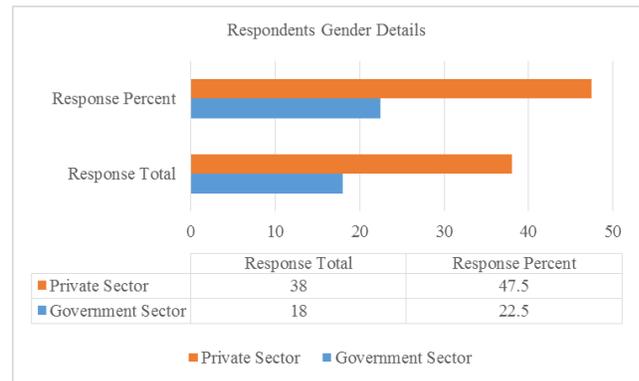

Fig. 1 Respondents Gender Details

Fig. 1 presents the gender profile of the respondents. The majority of the respondents were male which consist of 62.5%, and female consist of 37.5%. The corresponding mean for the age is 1.375, and the standard deviation is 0.484.

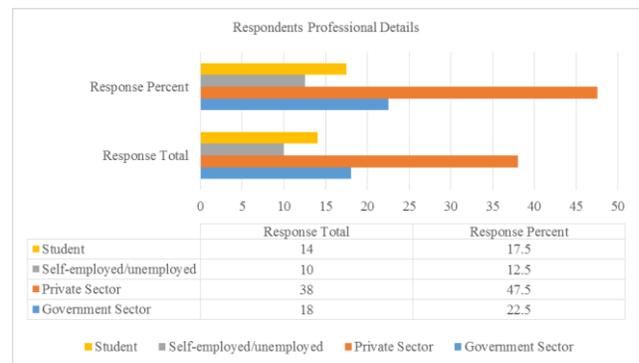

Fig. 2 Respondents Professional Details

Fig. 2 presents the profession profile of the respondents. Respondents that works in government sector consist 22.5%, 47.5% of the respondents' works in private sector, 12.5% of the respondents are either self-employed or unemployed, and 17.5% of the respondents are students the corresponding Mean and Standard Deviation are as follows, 2.25 and 0.994.

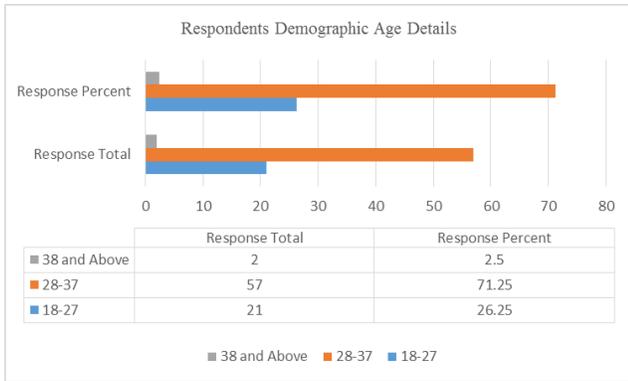

Fig. 3 Respondents Demographic Age Details

Fig.3 consists of Age profile of the respondents. The respondents between age brackets 28-37 was the highest responding to 71.25%, 18-27 age brackets consist of 26.25%, and matured respondents make up of 2.5%. The corresponding Means and Standard Deviation consist of 1.762 and 0.481 respectively.

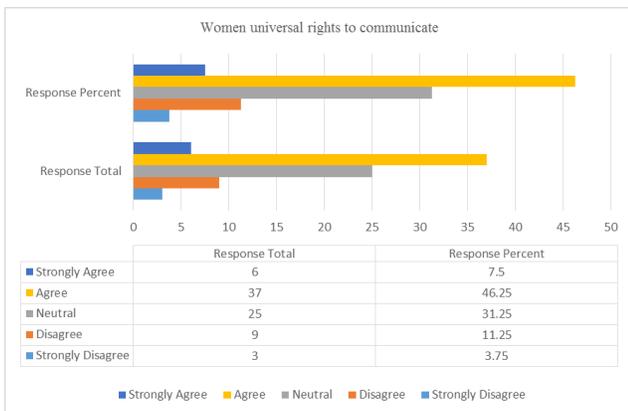

Fig. 4 Women universal rights to communicate

Fig.4 provides information on ICT usage habit, the respondents were requested to rate their opinion on a 5-point Likert scale; 5 for Strongly Disagree, 4 for Disagree, 3 for Neutral, 2 for Agree, and 1 for Strongly Agree. The results shows gender privilege of the right to communicate despite bias in communication connectivity. About 3.75% of respondents (Strongly Disagree) to the question, 11.25% (Disagree), and 31.25% of the respondents were (Neutral), while 46.25% of the respondents (Agree), and 7.5% (Strongly Agree) to the question. The corresponding Mean value and Standard Deviation are 3.425 and 0.919.

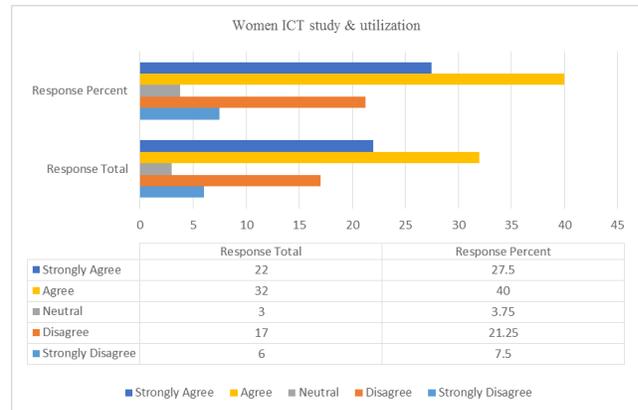

Fig. 5 Women ICT study & utilization

Fig.5 presents response on how less women study and use information technology than their men in Nigeria. About 7.5% of respondents (Strongly Disagree) to the question, 21.5% of the respondents (Disagree) that less women study and use ICT, 3.75% of the respondents remain (Neutral) to the question, while a high value of 40% (Agree) that less women study and use ICT, and 27.5% (Strongly Agree) with the fact that fewer women study and use ICT. The corresponding Mean value and Standard Deviation are 3.588 and 1.291. Analysis done using One-way Analysis of Variance showed that, it is significant that less women study and use information technology (F (1, 8) = 0.5290, P= 0.8225) at α = 0.05.

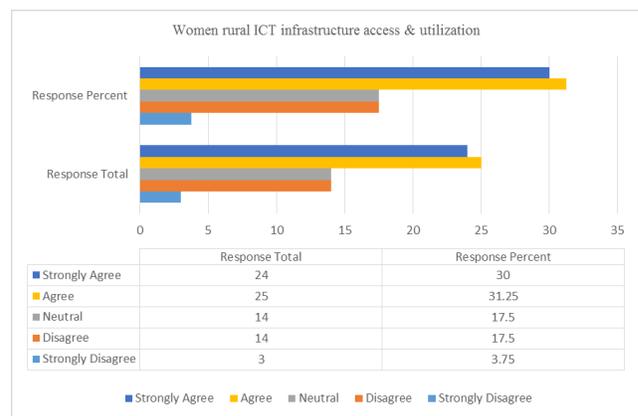

Fig. 6 Women rural ICT infrastructure access & utilization

Fig.6 highlighted how cultural, religious, and ethnic believes affects the use of information technology centers among women in rural areas in Nigeria. From the responses, it could be seen that there is cultural, religious and ethnic influence in using ICT facilities in rural Nigeria. About 3.75% of the respondents (Strongly Disagree) that less women use ICT facilities in rural area, 17.5% of the respondents (Disagree) that less women use ICT facilities in rural area, and another 17.5% of the respondents remain

(Neutral) to the question, while 31.25% of the respondents (Agree) that less women use ICT facilities in rural area, and 30% (Strongly Agree) with the fact that fewer women use ICT facilities in rural area. The corresponding Mean and Standard Deviation for the responses in Fig.6 are 3.588 and 1.291.

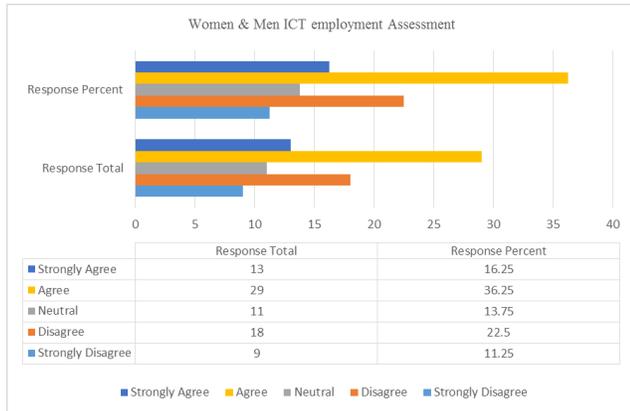

Fig. 7 Women & Men ICT employment Assessment

Fig.7 presents respondents response on women and men ICT employment assessment. About 11.25% of the respondents (Strongly Disagree) to the question, 22.5% of the respondents (Disagree) on the assertion that men tend to be highly productive in ICT employment than women, 13.75% were (Neutral) to the question, while 36.25% of the respondents (Agree) that men tend to be highly productive in ICT employment than women, and 16.25% (Strongly Agree) with the fact that men tend to be highly productive in ICT employment than women. The corresponding Mean value and Standard Deviation are 3.238 and 1.277 respectively. It implies that cultural and social attitude are responsible for women to be under productive than men, and allocation of more resources to men is also responsible for women not highly productive than men.

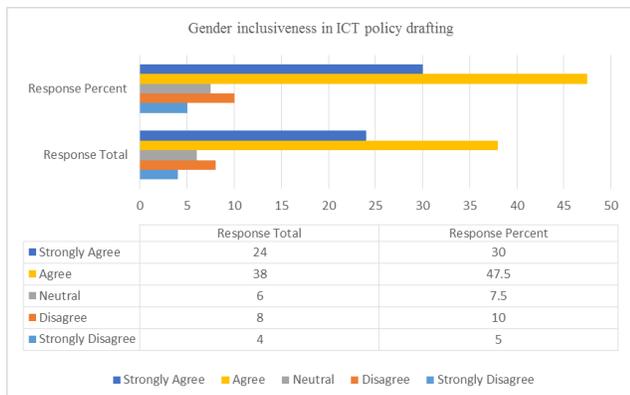

Fig. 8 Gender inclusiveness in ICT policy drafting

Fig.8 presents respondents response on Gender inclusiveness in ICT policy drafting. About 5% of the respondents (Strongly Disagree) to the question, 10% of the respondents (Disagree) on the assertion that Gender inclusiveness in ICT policy making will improve, and advance Nigeria information society and bring about development, 7.5% were (Neutral) to the question, while 47.5% of the respondents (Agree) that Gender inclusiveness in ICT policy making will improve, and advance Nigeria information society and bring about development, and 30% (Strongly Agree) with the fact that Gender inclusiveness in ICT policy making will improve, and advance Nigeria information society and bring about development. The corresponding Mean value and Standard Deviation are 3.911, and 1.058 respectively. This implies that development of Nigeria information technology society can be inclusive only if both men and women in the society share the benefit of developing Nigeria, to sustained social, psychological and economic progress.

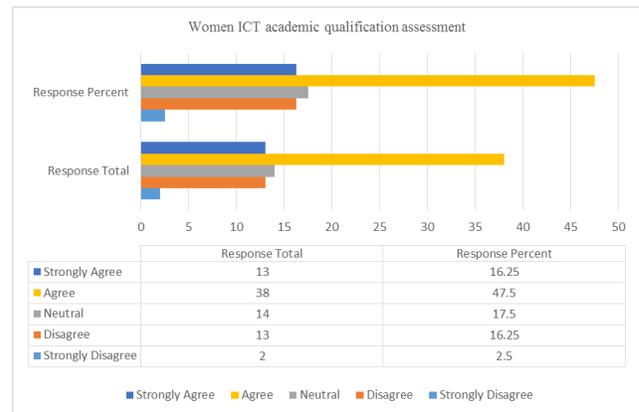

Fig. 9 Women higher ICT academic qualification assessment

Fig.9 presents respondents response on fewer women than men enroll in higher ICT academic qualification at the University. About 2.5% of the respondents (Strongly Disagree) to the question, 16.25% of the respondents (Disagree) on the assertion that fewer women than men enroll in ICT academic qualification at the University, 17.5% were (Neutral) to the question, while 47.5% of the respondents (Agree) that fewer women than men enroll in higher ICT academic qualification at the University, and 16.25% (Strongly Agree) with the fact that fewer women than men enroll in higher ICT academic qualification at the University. The corresponding Mean value and Standard Deviation are 3.556 and 1.054 respectively. This implies that Job roles plays a vital part for fewer women having ICT qualification as this is a psychological impact where balancing work and family life create less optimism in higher ICT academic qualification among women.

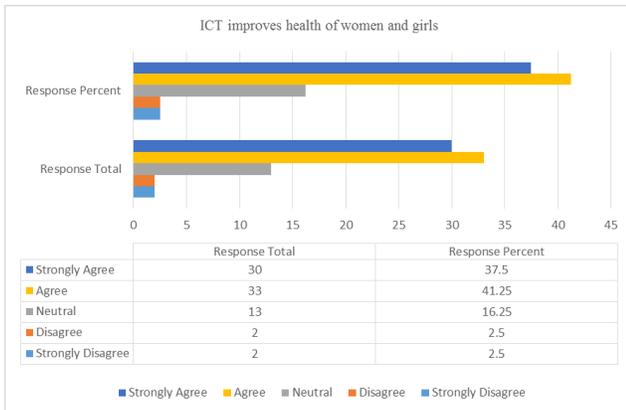

Fig. 10 ICT improves health of Women and Girls

Fig.10 highlight on how ICT can improve health situations of girls and women in Nigeria. From the survey, about 2.5% of the respondents (Strongly Disagree) to the question, 2.5% of the respondents (Disagree) on the assertion that ICT can improve health situation of girls and women in developing countries like Nigeria, 16.25% were (Neutral) to the question, while 41.25% of the respondents (Agree) that ICT can improve health situation of girls and women in Nigeria, and 37.5% (Strongly Agree) with the fact that ICT can improve health situation of girls and women in Nigeria.The corresponding mean and standard deviation for the respondents' response to the question are 4.127 and 0.862 respectively. From Fig.10, it implies that ICT has been useful in advancing gender health information including easy access to clinical data. It has drastically reduced emotional distressed about some certain health related diseases accessible on the world web.

## 5. Conclusion

Specific attention need to be emphasized to gender in order for Nigeria to achieve gender-positive result in ICT. According to Louise Chamberlain [32], he did stress the point that the need for gender analysis is to understand the difference in the outcome of integrating both men and women in policy design, and project implementation. This survey result has highlighted key gender issues that have been identified, and public opinion which should be of concern to Nigeria government and policy makers. A synergy between all stake-holders that would include gender equality should be encouraged and be sustained as this would advance ICT production in Nigeria. The need to work in coordination, collaboration, and corporation in ICT environment by all actors could yield a more inclusive, and democratically gender based information society in Nigeria. There is need for policy makers to identify and address fundamental issues like infrastructure, relevance, accessibility, affordability, and both computer software and hardware as stated in this research which relate to ICT and gender. In summary we recommend an all-inclusive deliberate ICT policy that attract and encourages women participation in ICT developmental framework.